\documentclass[12pt]{article}

\usepackage[utf8]{inputenc}
\usepackage{float}
\usepackage{authblk}
\usepackage{caption} 
\usepackage{subcaption}
\usepackage{cite}
\usepackage{setspace}
\usepackage{appendix}
\usepackage{makecell}
\usepackage[colorlinks=true,urlcolor=blue,citecolor=blue,linkcolor=red,linktocpage=true,pdfproducer=medialab]{hyperref}
\usepackage[a4paper,width=17cm,height=25cm]{geometry}
\usepackage{morefloats}
\usepackage{siunitx}
\usepackage{latexsym,amsfonts,amssymb,amsthm,amsmath}
\usepackage{amssymb,bm}
\usepackage{amsmath,graphicx,color,epsfig}
\usepackage{slashed}
\usepackage{hhline}
\usepackage{setspace}
\usepackage{braket}
\usepackage{xcolor}
\usepackage[dvipsnames]{xcolor}
\usepackage{fancyhdr}

\title{Vector Leptoquark Mediated Leptonic Decay of the Charged $B$-Meson}

\author[1]{M. Omar Nadeem\thanks{omarn791995@gmail.com}}
\author[2]{Arslan Sikandar\thanks{asikandar@phys.qau.edu.pk}}

\affil[1]{\textit{Department of Physics, School of Natural Sciences, National University of Sciences and Technology, Islamabad, Pakistan}}
\affil[2]{\textit{Department of Physics, Quaid-i-Azam University, Islamabad, Pakistan}}

\date{November 2025}

\begin{document}

\maketitle

\begin{abstract}
\noindent
We study the purely leptonic decay of the charged $B$-meson within the $U_1$ Vector Leptoquark model at both leading-order and with one-loop QCD corrections. The structure of this amplitude is characterised by direct quark-lepton couplings which allow for lepton flavour universality violation (LFUV) and additional loop-level topologies. The leptoquark channel contributions to the $B$-meson decay constant are computed and accommodated in its lattice-improved uncertainty bounds by constraining the parameter space of this model in two distinct new-physics scenarios. Under these constraints, we provide predictions for both the LFUV and non-LFUV contributions to the branching fractions of this decay. 
\end{abstract}

\section*{Introduction}
The decay constant is a non-perturbative parameter that quantifies the decay amplitude of mesons. Specifically, it is related to the hadronic part of the process; hence meson decay amplitudes and related observables can be parametrized in terms of the decay constant \cite{RevModPhys.88.035008,PAUL1998494}. Experimental observables are sensitive to new-physics effects like CP-violating phases and flavour-changing neutral currents; thus, determining the decay constant more precisely allows for better constraints on other parameters. \\
There are also strong indications of LFUV in the branching ratios of meson decays \cite{RevModPhys.94.015003,panda2024exploringleptonflavorviolating,sumensari2017leptonflavoruniversalityviolation}. While current experimental anomalies predominantly concern semi-leptonic \cite{Fleischer:2019wlx,Banelli:2018jad,PhysRevD.110.075004,guadagnoli2023leptonflavorviolationleptonflavoruniversalityviolation,RevModPhys.88.035008} decays, potential deviations from Standard Model (SM) predictions may also manifest in the decay widths and branching ratios of the closely related, yet theoretically simpler, purely leptonic decay channels \cite{guadagnoli2023leptonflavorviolationleptonflavoruniversalityviolation}. These depend on fewer experimental parameters and consequently, there are fewer sources of uncertainty. Any discrepancies due to LFUV can be accounted for by uncertainties in these parameters, among which the largest contribution is from the $B$-meson decay constant ($f_B$) which has an uncertainty of $\pm\,\:15\%$ in the Lattice QCD framework \cite{Aoki_2020}, corresponding to the tree-level process $B^{-}\to \ell\:\bar{\nu}_{\ell}$. Although QCD corrections are perturbative, they can be absorbed into the decay constant, giving an effective decay constant ($f_{B}^{eff}$) with an uncertainty of $\pm\,12\:\%$ with $\mathcal{O}(\alpha_s^3)$ in the framework of QCD Sum Rules \cite{Penin_2002}. The most modern estimates in the Lattice QCD framework at $\mathcal{O}(\alpha_s)$ \cite{Aoki_2020,Vaquero_Avil_s_Casco_2025,Na:2012kp} place the $B$-meson decay constant at about $f_{B}=0.19\:$GeV, with an uncertainty of $\pm\,2\:\%$.
New-physics effects can also be absorbed into $f_{B}^{eff}$ to account for the remaining uncertainty, which will constrain the new-physics parameters. There are currently no experimental indications of LFUV in the branching fractions of $B^- \to \ell_{i}^-\:\bar{\nu_{\ell_j}}$ but the branching fraction of $B^- \to \tau\:\bar{\nu_{\tau}}$ was measured by Belle \cite{Belle:2015odw} at around $(1.25\pm 0.28_{stat}\pm 0.27_{syst})\times 10^{-4}$ and a more recent measurement by Belle II \cite{gaudino2025mathcalbbrightarrowtaunumeasurementhadronicfei} at $(1.24\pm 0.41_{stat}\pm 0.19_{syst})\times 10^{-4}$, which still only corresponds to a $3\,\sigma$ deviation. The (non-LFUV) electron and muon modes are helicity suppressed; hence, no significant evidence exists for them. The upper limits of their branching fractions are $3.5\times 10^{-6}$ for the electron mode and $2.7\times 10^{-6}$ for the muon mode, at $90\:\%$ CL \cite{Belle:2014pdk}.
\bigbreak\noindent
In the SM, the purely leptonic decay of the $B^{-}$ meson proceeds via the point annihilation of the meson's quarks to a virtual weak boson and the subsequent production of a lepton and antineutrino of the corresponding flavour \cite{PAUL1998494,Briere:2023}. This does not admit LFUV at any order and the structure of the diagram only allows for vertex and external quark self-energy loops at leading-order in QCD. In recent times, both vector \cite{Ban_2022,PhysRevD.98.115002} and scalar \cite{Sahoo_2015,Yan_2019} leptoquark models have often been applied to anomalous effects associated with various semi-leptonic and rare decays of the $B$-meson. This paper deals with this process in a vector-leptoquark (VLQ) model which admits direct quark-lepton vertices \cite{Angelescu:2021lln,Korajac_2024}, thus avoiding a direct annihilation. The resulting tree-level topology allows for seven different one-loop QCD corrections. Due to two separate quark-lepton vertices, there are six LFUV channels, corresponding to all the different lepton--antineutrino flavour combinations, all of which contribute to the effective decay constant.
\bigbreak\noindent
The structure of this paper is organised as follows. We introduce the $U_1$ Vector Leptoquark Model in the first section and discuss its various quark-lepton couplings in detail. The second section establishes the theoretical framework of the purely leptonic $B^{-}$ decay amplitude within both the SM and the $U_1$ VLQ model, followed by a comparison that leads to an expression for the effective decay constant ($f_{B}^{eff}$) due to the VLQ amplitude. The third section improves upon these results with one-loop contributions to $f_{B}^{eff}$. The Numerical Analysis section discusses the graphical results depicting the behaviour of couplings, effective decay constants and branching fractions and explores an alternative new-physics scenario. The paper concludes with a summary of findings and a discussion of prospects in the Conclusion section. Important auxiliary results, including scalar coefficients from loop calculations and hadronic matrix elements, are provided in Appendices A and B, respectively.

\section*{$U_1$ Vector Leptoquark Model}
We work in the purely left-handed scenario of the $U_1$ Vector Leptoquark Model, which is the Standard Model extended with a massive $U_1$ vector field. The model preserves the SM gauge group, $\mathrm{SU}(3)_c \times \mathrm{SU}(2)_L \times \mathrm{U}(1)_Y$, with the $U_1$ VLQ transforming as $(3,1,2/3)$ \cite{Dorsner:2018ynv}. The model has the feature of allowing direct quark-lepton couplings. As the leptoquark is a colour triplet, its covariant derivative gives rise to both three-- and four--point gluon--leptoquark interaction terms with their vertex factors being similar to the Yang--Mills type vertices of vector-vector interactions, as given in \cite{shaw2026quantum}.\\
From the $U_1$ VLQ Lagrangian density given in \cite{Angelescu:2021lln,Korajac_2024}, we obtain the Hamiltonian density of the fermionic sector.

\begin{equation}
    \mathcal{H}_{U_1}(x)
= -x_{L}^{i\,j}\,\bar Q_i\,\gamma_\mu\,P_L\,L_j\,U_1^\mu+ \text{h.c} \dots
\end{equation}

Expanding out the fermion doublets reveals that up-type quarks couple to neutrinos while down-type quarks couple to charged leptons.

\begin{equation}
\mathcal{H}_{U_1}(x)
= -x_{L}^{i\,j}\,
\big(\bar u_{i}\,\gamma_\mu\,P_L\,\nu_{j}
+\bar d_{i}\,\gamma_\mu\,\ell_{j}\big)\,U_1^\mu +\text{h.c} \dots
\end{equation}

The first index ($i$) indicates the quark flavour and the second index ($j$) indicates the lepton flavour. It should be noted that the same combination of quark and lepton generations has the same coupling. This is of vital importance when considering the quark-neutrino interactions from the data given in \cite{Angelescu:2021lln}.

\section*{Tree Level Amplitude in $U_1$ Vector Leptoquark Model}

Purely leptonic decay amplitudes of heavy--light pseudoscalar mesons in the SM can be factorised in terms of their hadronic and leptonic matrix elements \cite{Grinstein2016}.

\begin{equation}
\mathcal{M}=\frac{G_F}{\sqrt{2}}\,V_{ub} \langle \ell \: \bar{\nu_{\ell}}|\bar{\ell}\gamma_{\mu}P_L\nu_{\ell}|0\rangle\,\langle 0|\bar{q}\gamma^{\mu}P_L\:b|B\rangle \;.
\end{equation}

This proceeds via the annihilation of the $b$ and light quarks into a virtual weak boson and the subsequent production of a lepton and antineutrino. These matrix elements can be parametrized as follows:

\begin{equation}
\langle \ell \: \bar{\nu_{\ell}}|\bar{\ell}\gamma_{\mu}P_L\nu_{\ell}|0\rangle=\bar{u}_l\,\gamma_{\mu}\,P_L\,\nu_{\nu_{\ell}} \;,
\end{equation}

\begin{equation}
\langle 0|\bar{q}\gamma^{\mu}P_L\:b|B\rangle=i\,f_B\,p^{\mu} \;.
\end{equation}

The hadronic matrix element can be parametrized in this way because of the direct annihilation of the quarks in the meson. However, the heavy virtual mediator between the two vertices can be integrated out, yielding a four-point topology that directly couples the two quarks and allows the amplitude to be parametrized in terms of the decay constant.

\begin{figure}[h!]
    \centering
    \includegraphics[width=0.5\linewidth]{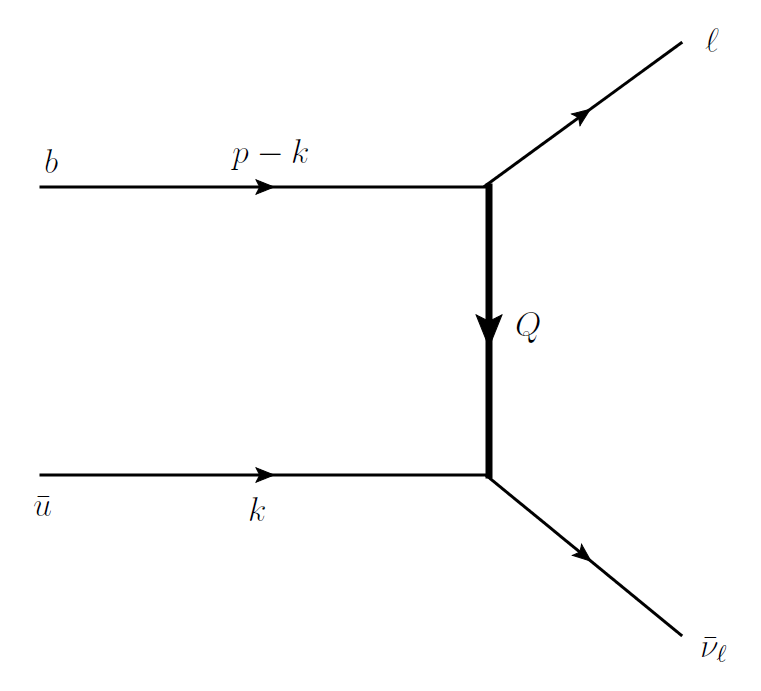}
    \caption{Tree level diagram of the $U_1$ vector Leptoquark mediated decay of the $B$-meson. This is the leading order amplitude of this process in this model.}
    \label{fig:tree}
\end{figure}

Considering a general case of $B^- \to \ell_{i}^-\:\bar{\nu_{\ell_j}}$, the tree-level leptoquark mediated amplitude (Fig.\ref{fig:tree}) is

\begin{equation}
i\,\mathcal{M}^{(U_1,\,0)}=\bar{u}_{\ell}\:(i\,x_{b\ell}\gamma^{\beta}P_L)\,u_b\frac{-i\,g_{\beta\lambda}}{M^2}\bar{\nu}_{u}\:(i\,x_{u\nu}\gamma^{\lambda}P_L)\:\nu_{\nu_{\ell}} \;.
\end{equation}

After approximating this amplitude as a four-point interaction and applying a Fierz transformation to interchange the spinors, its effective Hamiltonian density is

\begin{equation}
\mathcal{H}_{U_1}^{\rm eff}=\frac{1}{M^2}\,(x_{L}^{3\,i}\,x_{L}^{1\,j})\, (\bar{u}\,\gamma^{\mu}\,P_L\, b)\,(\bar{\ell_i}\,\gamma_{\mu}\,P_L\,\nu_{\ell_j}) \;.
\end{equation}

This makes it proportional to the SM effective Hamiltonian for this process and hence the product of bilinears can be represented by SM-like matrix elements,

\begin{equation}
\mathcal{M}_{U_1}^{(0)}=\frac{1}{M^2}\,(x_{L}^{3\,i}\,x_{L}^{1\,j})\, \langle \bar{\ell_i}\,\nu_{\ell_j}|\ell_i\,\gamma_{\mu}\,P_L\,\bar{\nu}_{\ell_j} |0\rangle \, \langle 0 | \bar{u}\,\gamma^{\mu}\,P_L\,b |B\rangle \;,
\end{equation}

\begin{equation}
\mathcal{M}_{U_1}^{(0)}=\frac{1}{M^2}\,(x_{L}^{3\,i}\,x_{L}^{1\,j})\, (\bar{u}_{\ell_i}\,\gamma_{\mu}\,P_L\,\nu_{\ell_j})\, (i\,f_{B}^{(0)}\,p^{\mu}) \;.
\end{equation}

To compare this with the amplitude of the same process in the SM, the $U_1$ VLQ-mediated amplitude can be cast into a form similar to the SM-mediated amplitude by absorbing the extra parameters into the leading order SM decay constant as follows:

$$\mathcal{M}_{U_1}^{(0)}=\frac{G_F}{\sqrt{2}}\,V_{ub}\, (\bar{u}_{\ell_i}\,\gamma_{\mu}\,P_L\,\nu_{\ell_j})\, (i\,f_{B}^{i\,j\,(0)}\,p^{\mu})\;.$$

Hence, the effective decay constant for the leptoquark mediated process at leading order is given as follows, where the general relation is, for $i$ lepton flavour and $j$ antineutrino flavour,

$$f_{B}^{i\,j\,(0)}=\left(\frac{x_{L}^{3\,i}\,x_{L}^{1\,j}\,\sqrt{2}}{G_F\,V_{ub}\,M^2}\right)\,f_{B}^{(0)}\;.$$
Here the first index indicates lepton flavour and the second indicates neutrino flavour. For $i\,\neq\,j$, there are decay constant contributions from LFU-violating processes as well. \\
Even with the four-point effective interaction, the SM does not admit LFU-violating terms and lepton flavour plays no role in the leading order SM contribution to the meson decay constant.

\section*{One Loop Amplitudes in the $U_1$ Vector Leptoquark Model}

All of the following one-loop amplitudes have been computed using the package \cite{Patel_2017} and all Feynman diagrams have been drawn using the software JaxoDraw \cite{Binosi_2004}. The results of the loop integrals presented here have been evaluated at $\mu=m_b$ and have been renormalized with the modified minimal subtraction scheme and we have retained only the kinematically leading-order terms of each integral.

\subsection*{$b$ Quark Vertex Loop}

\begin{figure}[h]
    \centering
    \includegraphics[width=0.5\linewidth]{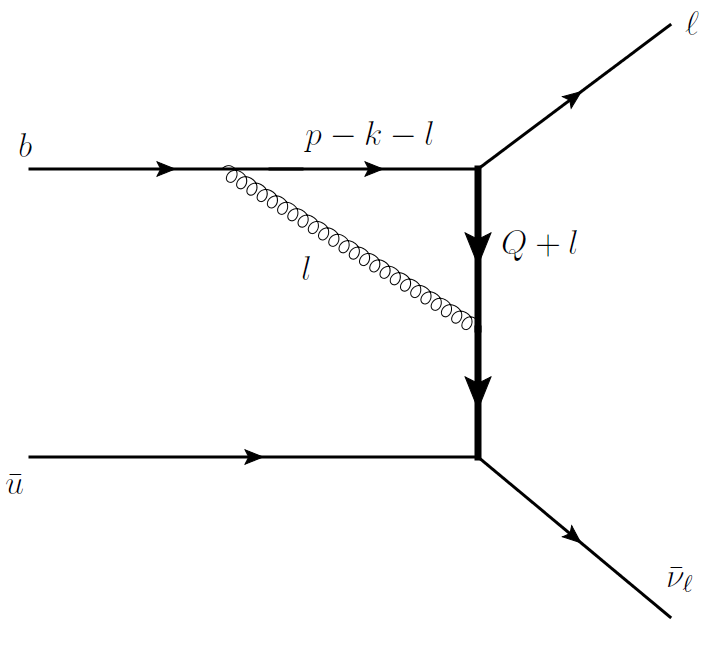}
    \caption{Leptoquark-mediated decay of the $B$-meson, with a $b$ quark vertex correction.}
    \label{fig:Vertexbl}
\end{figure}

From Fig.\ref{fig:Vertexbl},

\begin{equation}
i\,\mathcal{M}_{b}^{(1)}=\bar{u}_{\ell}\:(x_{b\ell}\,\Gamma_{b}^{\rho})\:u_b\frac{-i\,g_{\rho\lambda}}{M^2}\bar{\nu}_{u}\:(i\,x_{u\nu}\gamma^{\lambda}P_L)\,\nu_{\nu_{\ell}} \;,
\end{equation}

\begin{equation}
\begin{split}
        \Gamma_{b}^{\rho}=&\mu^{2\epsilon}\,C_F\,\int\frac{d^Dl}{(2\,\pi)^D}(-i\,g\gamma^{\alpha})\frac{i\,(\slashed{p}-\slashed{k}-\slashed{l}+m_b)}{(p-k-l)^2-m_{b}^{2}}\frac{-i\,g_{\alpha\mu}}{l^2} \, (i\,\gamma^{\beta}P_L)\frac{-i\,g_{\beta\nu}}{M^2}\left(\frac{2}{3}\,i\,g\right) \\
        & \times [g^{\mu\nu}\:(2\,l-Q)^{\rho}+g^{\nu\rho}\:(2\,Q-l)^{\mu}+g^{\rho\mu}\:(-Q-l)^{\nu}] \;,
\end{split}
\end{equation}

\begin{equation}
\Gamma_{b}^{\rho} = \frac{g^{2}\,P_{L}}{M^{2}}\Big(a_{1}\,p^{\rho}+ a_{2}\,\slashed{p}\,p^{\rho}
+ a_{3}\,p_{\ell}^{\rho}+ a_{4}\,\slashed{p}\,p_{\ell}^{\rho}+ a_{5}\,\gamma^{\rho}+ a_{6}\,\slashed{p}\,\gamma^{\rho}+ a_{7}\,\slashed{p}_{\ell}\,\gamma^{\rho}+a_{8}\,\slashed{p}\,\slashed{p}_{\ell}\,\gamma^{\rho}
\Big) \;.
\end{equation}

After substituting this factored expression of the loop structure into the full one-loop amplitude, each of the resulting $b$-quark/lepton bilinears have a different Dirac structure. Similar to the tree-level case, after the Fierz transformation the amplitude can be factored into  hadronic and leptonic matrix elements. 

\begin{equation}
     i\:\mathcal{M}_{b}^{(1)}=\frac{g^2\:(x_{L}^{3\,i}\,x_{L}^{1\,j})}{M^4}\: \langle \bar{\ell_i}\,\nu_{\ell_j}|\ell_i\,\gamma_{\mu}\,P_L\,\bar{\nu}_{\ell_j} |0\rangle \:H_{b}^{\mu} \;,
\end{equation}

where

\begin{equation}
        H_{b}^{\mu} = \langle 0 | \bar{u}\,P_L\,(a_1\,p^{\mu}+a_2\,\slashed{p}\,p^{\mu}+a_3\,p_{\ell}^{\mu}+a_4\,\slashed{p}\,p_{\ell}^{\mu}+a_5\,\gamma^{\mu}+ a_6\,\slashed{p}\,\gamma^{\mu} +a_7\,\slashed{p}_{\ell}\,\gamma^{\mu}+ a_8\,\slashed{p}\,\slashed{p}_{\ell}\,\gamma^{\mu})\,b |B\rangle \;.
\end{equation}

Simplifying the hadronic matrix element

\begin{equation}
H_{b}^{\mu} 
= -i\,f_{B}^{(0)}\,\left(A\,p^{\mu} + B\,p_{\ell}^{\mu}\right) \;,
\end{equation}

\begin{equation}
\noindent
\begin{minipage}{0.45\textwidth}
$$
A = \frac{m_B^{2}}{m_b+m}\,(a_1+a_6) + m_B^{2}\,a_2 + a_5 \;,
$$
\end{minipage}
\hfill
\begin{minipage}{0.45\textwidth}
$$
B = \frac{m_B^{2}}{m_b+m}\,(a_3+a_7) + m_B^{2}\,(a_4+a_8) \;.
$$
\end{minipage}
\end{equation}

Hence,

\begin{equation}
\mathcal{M}_{b}^{(1)}= \frac{g^2\,(x_{L}^{3\,i}\,x_{L}^{1\,j})}{M^4}\:\langle \bar{\ell_i}\,\nu_{\ell_j}|\ell_i\,\gamma_{\mu}\,P_L\,\bar{\nu}_{\ell_j} |0\rangle \times \Bigg[-i\,f_{B}^{(0)}\,\left(A\,p^{\mu} + B\,p_{\ell}^{\mu}\right)\Bigg] \;.
\end{equation}

After applying the equation of motion to the spinors and the SM normalization and absorbing the extra factors into $f_{B}^{(0)}$,

\begin{equation}
\mathcal{M}_{b}^{(1)} =\frac{G_F\,V_{ub}}{\sqrt{2}}\,\bar{u}_{\ell_i}\,P_L\,\nu_{\ell_j} \,(i\,m_{\ell_i}\,f_{B}^{i\,j\,(b)}) \;.
\end{equation}

Thus, the effective decay constant with the $U_1$ VLQ-mediated amplitude, with a gluonic correction on the $b$ quark vertex, is

\begin{equation}
f_{B}^{i\,j\,(b)}=-\frac{g^2\,(x_{L}^{3\,i}\,x_{L}^{1\,j})}{M^4}\:\left(\frac{A+B}{G_F\,V_{ub}/\sqrt{2}}\right)\:f_{B}^{(0)}\;.
\end{equation}

\pagebreak

\subsection*{$\bar{u}$ Quark Vertex Loop}

\begin{figure}[h]
    \centering
    \includegraphics[width=0.5\linewidth]{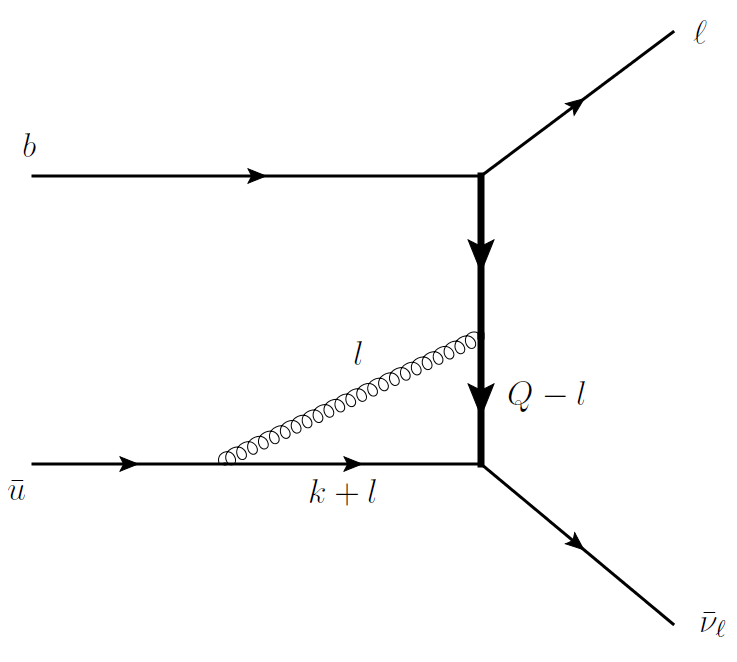}
    \caption{Leptoquark-mediated decay of $B$-meson, with a light quark vertex correction.}
    \label{fig:vertexuv}
\end{figure}

From Fig.\ref{fig:vertexuv},

\begin{equation}
i\,\mathcal{M}_{u}^{(1)}=\bar{u}_{\ell}\:(i\,x_{b\ell}\,\gamma^{\beta}P_L)\:u_b\frac{-i\,g_{\beta\mu}}{M^2}\bar{\nu}_{u}\:(x_{u\nu}\,\Gamma_{u}^{\mu})\,\nu_{\nu_{\ell}} \;,
\end{equation}

\begin{equation}
\Gamma_{u}^{\mu} =
\frac{g^{2}\,P_{L}}{M^{2}}\Big(b_{1}\,p^{\mu}+ b_{2}\,\slashed{k}\,p^{\mu}+ b_{3}\,p_{\ell}^{\mu}
+ b_{4}\,\slashed{k}\,p_{\ell}^{\mu}+ b_{5}\,\slashed{p}\,\gamma^{\mu}+ b_{6}\,\slashed{k}\,\slashed{p}\,\gamma^{\mu}+ b_{7}\,\slashed{p}_{\ell}\,\gamma^{\mu}+ b_{8}\,\slashed{k}\,\slashed{p}_{\ell}\,\gamma^{\mu} 
\Big) \;.
\end{equation}

Following a similar procedure to that used for the previous vertex loop, we obtain

\begin{equation}
\mathcal{M}_{u}^{(1)}= \frac{g^2\,(x_{L}^{3\,i}\,x_{L}^{1\,j})}{M^4}\:\langle \bar{\ell_i}\,\nu_{\ell_j}|\ell_i\,\gamma_{\mu}\,P_L\,\bar{\nu}_{\ell_j} |0\rangle \times \Bigg[-i\,f_{B}^{(0)}\,\left(C\,p^{\mu} + D\,p_{\ell}^{\mu}\right)\Bigg] \;,
\end{equation}

Thus, its contribution to the effective decay constant is 

\begin{equation}
f_{B}^{i\,j\,(u)}=-\frac{g^2\,(x_{L}^{3\,i}\,x_{L}^{1\,j})}{M^4}\:\left(\frac{m_B^{2}}{m_b+m}\,\frac{C+D}{G_F\,V_{ub}/\sqrt{2}}\right)\:f_{B}^{(0)} \;,
\end{equation}

where

\begin{equation}
\noindent
\begin{minipage}{0.45\textwidth}
$$
C = \frac{m_{B}^{2}}{m_{b}+m}\,\Big(\,b_{1}+b_{5}+m\,(b_{2}+b_{6})\,\Big) \;,
$$
\end{minipage}
\hfill
\begin{minipage}{0.45\textwidth}
$$
D = \frac{m_{B}^{2}}{m_{b}+m}\,\Big(\,b_{3}+b_{7}+m\,(b_{4}+b_{8})\,\Big) \;.
$$
\end{minipage}
\end{equation}

\subsection*{Vector Leptoquark Self Energy}

\begin{figure}[h!]
    \centering
    \begin{subfigure}[b]{0.45\linewidth}
        \centering
        \includegraphics[width=\linewidth]{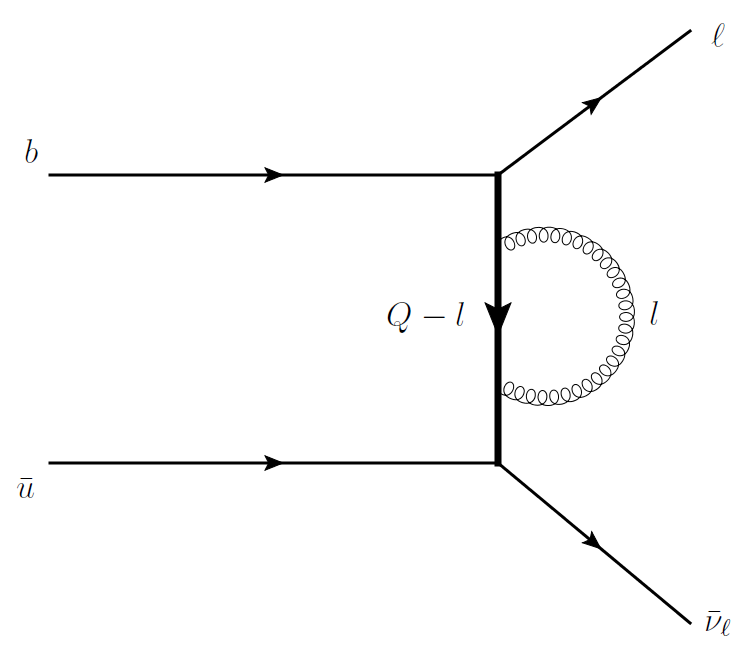}
        \caption{$U_1$ VLQ-mediated $B$ decay with three-point leptoquark self-energy correction.}
        \label{fig:se3}
    \end{subfigure}
    \hfill
    \begin{subfigure}[b]{0.45\linewidth}
        \centering
        \includegraphics[width=\linewidth]{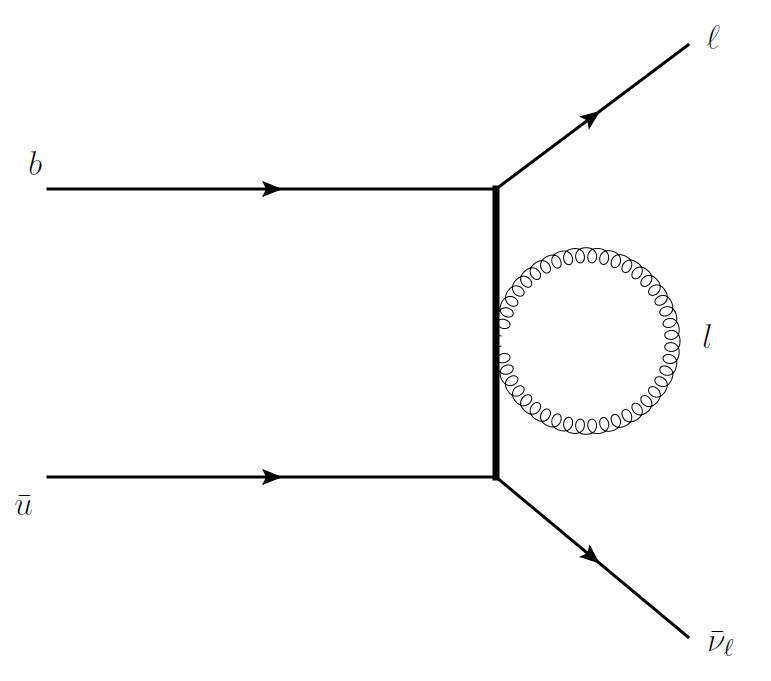}
        \caption{$U_1$ VLQ-mediated $B$ decay with four-point leptoquark self-energy correction.}
        \label{fig:se4}
    \end{subfigure}
    \caption{Gluon self-energy loops on the internal VLQ.}
    \label{fig:side_by_side_se}
\end{figure}

There are two different leptoquark self-energy diagrams due to the three-point (Fig.\ref{fig:se3}) and four-point vertices (Fig.\ref{fig:se4}) of leptoquarks with gluons.

The amplitude with the three-point gluon-leptoquark self-energy is given as follows,

\begin{equation}
i\,\mathcal{M}_{\Sigma_{U_1}}^{(1)}=\bar{u}_{\ell}\:(i\,x_{b\ell}\gamma^{\beta}P_L)\:u_b\frac{-i\,g_{\beta\sigma}}{M^2}(i\,\Sigma_{U_1}^{\sigma\mu})\frac{-i\,g_{\mu\lambda}}{M^2}\bar{\nu}_{u}\:(i\,x_{u\nu}\gamma^{\lambda}P_L)\:\nu_{\nu_{\ell}} \;.
\end{equation}

After contracting the Lorentz indices, the amplitude becomes

\begin{equation}
i\,\mathcal{M}_{\Sigma_{U_1}}^{(1)}=\bar{u}_{\ell}\:(i\,x_{b\ell}\gamma^{\beta}P_L)\:u_b\:\frac{i\,\Sigma_{\beta\lambda}^{U_1}}{M^4}\,\bar{\nu}_{u}\:(i\,x_{u\nu}\gamma^{\lambda}P_L)\:\nu_{\nu_{\ell}} \;,
\end{equation}

where

\begin{equation}
    \begin{split}
        i\,\Sigma_{U_1}^{\beta\lambda}=&g^{2}\,
\Bigg(
 \frac{10\,M^{4}}{27\,(k \!\cdot\! p_{\ell})^{2}}\,p^{\lambda}\,p^{\beta}
 -\frac{20\,M^{2}}{27\,(k \!\cdot\! p_{\ell})}\,p^{\lambda}\,p^{\beta}
 -\frac{10\,M^{4}}{27\,(k \!\cdot\! p_{\ell})^{2}}\,p^{\beta}\,p_{\ell}^{\lambda}\\
 &+\frac{20\,M^{2}}{27\,(k \!\cdot\! p_{\ell})}\,p^{\beta}\,p_{\ell}^{\lambda}
 -\frac{10\,M^{4}}{27\,(k \!\cdot\! p_{\ell})^{2}}\,p^{\lambda}\,p_{\ell}^{\beta}
 +\frac{20\,M^{2}}{27\,(k \!\cdot\! p_{\ell})}\,p^{\lambda}\,p_{\ell}^{\beta}\\
& +\frac{10\,M^{4}}{27\,(k \!\cdot\! p_{\ell})^{2}}\,p_{\ell}^{\lambda}\,p_{\ell}^{\beta}
 -\frac{20\,M^{2}}{27\,(k \!\cdot\! p_{\ell})}\,p_{\ell}^{\lambda}\,p_{\ell}^{\beta}
\Bigg) \;.
    \end{split}
\end{equation}

Assuming a point interaction of the quarks and leptons, the loop structure and amplitude can be modified as follows

\begin{equation}
i\,\mathcal{M}_{\Sigma_{U_1}}^{(1)}=\bar{u}_{\ell}\:(i\,x_{b\ell}\gamma^{\beta}P_L)\:u_b\:\frac{i\,\Sigma^{U_1}}{M^4}\,\bar{\nu}_{u}\:(i\,x_{u\nu}\gamma_{\beta}P_L)\:\nu_{\nu_{\ell}} \;,
\end{equation}

where 

\begin{equation}
    \begin{split}
        i\,\Sigma_{U_1} =& g^{2}\,
\Bigg(
 \frac{10\,M^{4}}{27\,(k \!\cdot\! p_{\ell})^{2}}\,p^2
 -\frac{20\,M^{2}}{27\,(k \!\cdot\! p_{\ell})}\,p^2
 -\frac{10\,M^{4}}{27\,(k \!\cdot\! p_{\ell})^{2}}\,p.p_{\ell}\\
 &+\frac{20\,M^{2}}{27\,(k \!\cdot\! p_{\ell})}\,p.p_{\ell}
 -\frac{10\,M^{4}}{27\,(k \!\cdot\! p_{\ell})^{2}}\,p.p_{\ell}
 +\frac{20\,M^{2}}{27\,(k \!\cdot\! p_{\ell})}\,p.p_{\ell}\\
& +\frac{10\,M^{4}}{27\,(k \!\cdot\! p_{\ell})^{2}}\,p_{\ell}^{2}
 -\frac{20\,M^{2}}{27\,(k \!\cdot\! p_{\ell})}\,p_{\ell}^{2}
\Bigg) \;.
    \end{split}
\end{equation}

Simplifying the above,

\begin{equation}
\begin{split}
i\,\Sigma_{U_1} =& \,g^{2}\,
 \frac{10}{27}\Bigg[
\frac{M^{4}}{(k\!\cdot\!p_{\ell})^{2}}\,m_{B}^{2}
-\frac{2M^{4}}{(k\!\cdot\!p_{\ell})^{2}}\,(p\!\cdot\!p_{\ell})
+\frac{M^{4}}{(k\!\cdot\!p_{\ell})^{2}}\,m_{\ell}^{2} \\
& -\frac{2M^{2}}{(k\!\cdot\!p_{\ell})}\,m_{B}^{2}
+\frac{4M^{2}}{(k\!\cdot\!p_{\ell})}\,(p\!\cdot\!p_{\ell})
-\frac{2M^{2}}{(k\!\cdot\!p_{\ell})}\,m_{\ell}^{2}
\Bigg]\;.
\end{split}
\end{equation}

If this diagram is considered as a point interaction, then due to the kinematics of the process

\begin{equation}
p.p_{\ell}=\frac{m_{B}^{2}+m_{\ell}^{2}}{2\:m_B}\;,
\end{equation}

\begin{equation}
i\,\Sigma_{U_1}=0\;.
\end{equation}

Hence, there is no contribution to the decay constant from either of the vector leptoquark self-energy diagrams.

\pagebreak

\subsection*{Hadronic Box}

\begin{figure}[h!]
    \centering
    \includegraphics[width=0.5\linewidth]{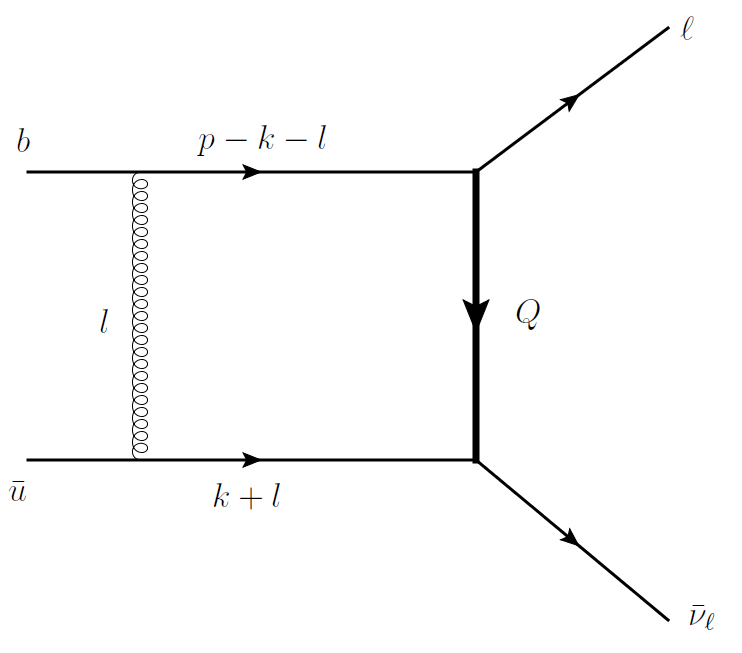}
    \caption{$b$ and $\bar{u}$ quarks exchanging a virtual gluon to form a hadronic box diagram.}
    \label{fig:boxh}
\end{figure}

From Fig.\ref{fig:boxh},

\begin{equation}
i\,\mathcal{M}_{Box}^{(1)}=\bar{u}_{\ell}\:(i\,x_{b\ell}\:\gamma^{\mu}P_L)\:u_b\:\frac{-i\,g_{\mu\nu}}{M^2}\,\beta\:\bar{\nu}_{u}\:(i\,x_{u\nu}\:\gamma^{\nu}P_L)\:\nu_{\nu_{\ell}} \;,
\end{equation}

\begin{equation}
\beta=i\,g^2\,\Big(c_{1}\,\mathbb{I}+ c_{2}\,\slashed{p}\Big) \;.
\end{equation}

Following a similar procedure as for the vertex loops,

\begin{equation}
\mathcal{M}_{Box}^{(1)}
= \frac{g^2\,(x_{L}^{3\,i}\,x_{L}^{1\,j})}{M^4}\:\langle\bar{\ell_i}\,\nu_{\ell_j}|\ell_i\,\gamma_{\mu}\,P_L\,\bar{\nu}_{\ell_j} |0\rangle \times \Bigg[i\,f_{B}^{(0)}\,p^{\mu}\,E\Bigg] \;.
\end{equation}

Thus, the effective decay constant in the $U_1$ VLQ model, with a hadronic box correction, is

\begin{equation}
f_{B}^{i\,j\,(Box)}=\frac{g^2\,(x_{L}^{3\,i}\,x_{L}^{1\,j})}{M^4}\,\:\left(\frac{E}{G_F\,V_{ub}/\sqrt{2}}\right)\:f_{B}^{(0)} \;.
\end{equation}

where

\begin{equation}
E=c_2\,\frac{m_B^{2}}{m_b+m} -c_1\;.
\end{equation}

\pagebreak

\subsection*{External Quark Self Energy}

\begin{figure}[h!]
    \centering
    \begin{subfigure}[b]{0.45\linewidth}
        \centering
        \includegraphics[width=\linewidth]{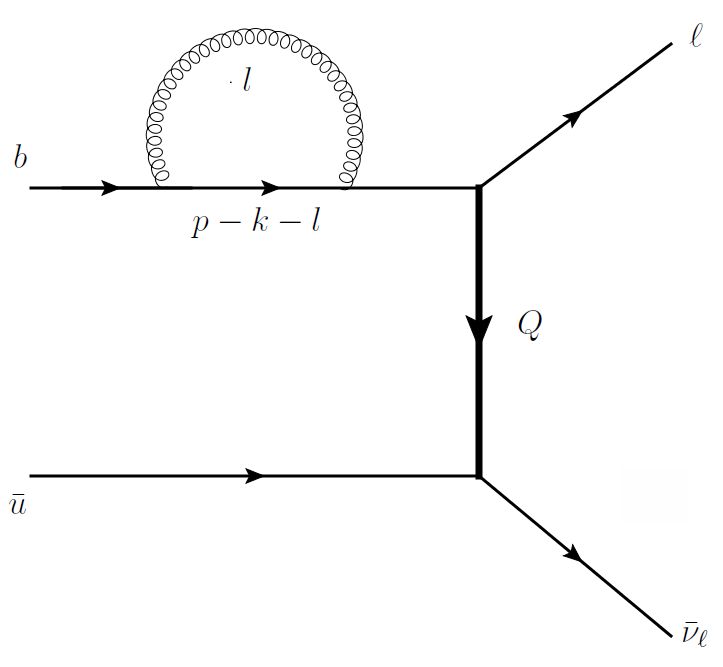}
        \caption{External $b$ quark self-energy diagram.}
        \label{fig:seb}
    \end{subfigure}
    \hfill
    \begin{subfigure}[b]{0.45\linewidth}
        \centering
        \includegraphics[width=\linewidth]{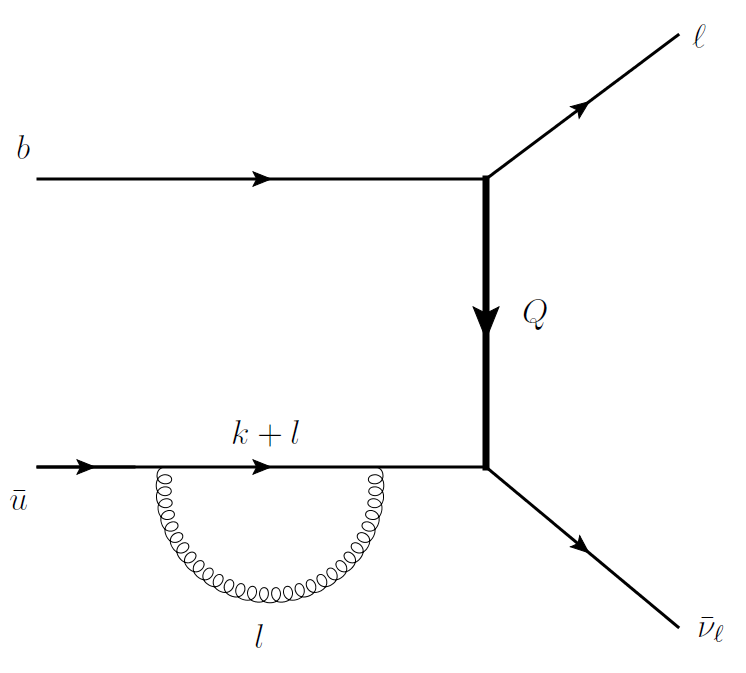}
        \caption{External $\bar{u}$ quark self-energy diagram.}
        \label{fig:seu}
    \end{subfigure}
    \caption{Gluon self-energy loop on the external quarks.}
    \label{fig:side_by_side}
\end{figure}

Following the approach of \cite{Descotes_Genon_2003} for the external quark loop corrections, the external heavy quark (Fig.\ref{fig:seb}) and light quark self-energy correction amplitudes (Fig.\ref{fig:seu}) are given as

\begin{equation}
\mathcal{M}_{\Sigma_b}^{(1)}=\frac{\delta_b}{2}\:\mathcal{M}^{(0)}\;\;\text{and}\;\; 
\mathcal{M}_{\Sigma_{\bar{u}}}^{(1)}=\frac{\delta_{\bar{u}}}{2}\,\mathcal{M}^{(0)}\;,
\end{equation}

where 

\begin{equation}
    \delta_b=i\,\frac{d\,\Sigma_b}{d\,\slashed{p}}\,\Bigg|_{\slashed{p}=m_b} \;\;\text{and}\;\; \delta_{\bar{u}}=i\,\frac{d\,\Sigma_{\bar{u}}}{d\,\slashed{k}}\,\Bigg|_{\slashed{k}=m} \;.
\end{equation}

As both external-leg self-energy amplitudes are proportional to the tree-level amplitude, after Fierz transformations and SM normalizations their parameterization is the same as that of the tree-level amplitude but with a scalar prefactor.

\begin{equation}
\mathcal{M}_{U_1}^{(0)}=\frac{1}{M^2}\,(x_{L}^{3\,i}\,x_{L}^{1\,j})\, (\bar{u}_{\ell_i}\,\gamma_{\mu}\,P_L\,\nu_{\ell_j})\,\left(i\,f_{B}^{(0)}\,\frac{\delta_{b,\bar{u}}}{2}\,p^{\mu}\right) \;.
\end{equation}

Comparing this with the SM amplitude and applying the respective spinor's equation of motion gives the effective decay constants in the $U_1$ VLQ model, with external quark leg self-energy corrections

\begin{equation}
f_{B}^{i\,j\,(\Sigma)}=\frac{g^2\,(x_{L}^{3\,i}\,x_{L}^{1\,j})}{M^4}\,\:\left(\frac{\tilde{\delta}_{b,\bar{u}}/2}{G_F\,V_{ub}/\sqrt{2}}\right)\:f_{B}^{(0)}\;.
\end{equation}

\section*{Numerical Analysis}

In this section we present the plots of quark-lepton couplings, the $B$-meson effective decay constants and the branching fractions, against the predicted leptoquark masses. The uncertainty of the decay constant is used to constrain the couplings, and we then present the branching fractions for both LFUV and non-LFUV channels.

\subsubsection*{Decay Constant and Couplings}

Table 3 of \cite{Angelescu:2021lln} gives the upper limits of the couplings of second and third generation quark and leptons as functions of the VLQ mass. These limits were obtained from indirect searches for Leptoquarks from LHC experiments involving dilepton production from proton--proton collisions \cite{CMS:2019tbu,2020PhRvL.125e1801A}. The couplings involving first generation quarks and leptons were approximated from the given trends by fitting with the power law regression model. This follows from the observation that the upper limits of the second and third generation fermion couplings in the $U_1$ VLQ model are approximately linear functions of the leptoquark mass for values above 2\:TeV. The slopes and vertical intercepts of those graphs tend to increase with quark and lepton masses.

\setlength{\tabcolsep}{4pt} 
\renewcommand{\arraystretch}{1.5} 
\begin{table}[h!]
\centering
\begin{tabular}{|c|c|c|c|c|c|}
\hline
\textbf{Coupling} & $a$ & $b$ & $m_q$/GeV & $m_{\ell}$/GeV & \textbf{Equation of the Line} \\
\hline
$x_{L}^{2\,3}$ & $5.0\times10^{-4}$  & 0.20  & $m_s = 0.095$ & $m_{\tau} = 1.777$ & $x_{L}^{2\,3} = (5.0\times10^{-4})\,M + 0.20$ \\
$x_{L}^{2\,2}$ & $3.75\times10^{-4}$ & 0.20  & $m_s = 0.095$ & $m_{\mu} = 0.106$ & $x_{L}^{2\,2} = (3.75\times10^{-4})\,M + 0.20$ \\
$x_{L}^{3\,3}$ & $7.5\times10^{-4}$  & 0.20  & $m_b = 4.18$  & $m_{\tau} = 1.777$ & $x_{L}^{3\,3}= (7.5\times10^{-4})\,M + 0.20$ \\
$x_{L}^{3\,2}$ & $5.5\times10^{-4}$  & 0.30  & $m_b = 4.18$  & $m_{\mu} = 0.106$ & $x_{L}^{3\,2} = (5.5\times10^{-4})\,M + 0.30$ \\
\hline
$x_{L}^{3\,1}$ & $3.13\times10^{-4}$ & 0.398 & $m_b = 4.18$  & $m_e = 0.0005$ & $x_{L}^{3\,1} = (3.13\times10^{-4})\,M + 0.398$ \\
$x_{L}^{1\,1}$ & $1.55\times10^{-4}$ & 0.278 & $m_d = 0.005$ & $m_e = 0.0005$ & $x_{L}^{1\,1} = (1.55\times10^{-4})\,M + 0.278$ \\
$x_{L}^{1\,2}$ & $2.74\times10^{-4}$ & 0.189 & $m_d = 0.005$ & $m_{\mu} = 0.106$ & $x_{L}^{1\,2} = (2.74\times10^{-4})\,M + 0.189$ \\
$x_{L}^{1\,3}$ & $3.70\times10^{-4}$ & 0.154 & $m_d = 0.005$ & $m_{\tau} = 1.777$ & $x_{L}^{1\,3} = (3.70\times10^{-4})\,M + 0.154$ \\
\hline
\end{tabular}
\caption{Upper limits of the coupling parameters $x_L^{ij}$, showing their dependence on the leptoquark mass as linear functions with corresponding quark and lepton masses.}
\label{tab:couplings}
\end{table}

The quark-lepton couplings in Table.\ref{tab:couplings} are given by the following equation:

\begin{equation}
    x_{L}^{i\,j}=a\,M+b \;,
\end{equation}

where

\begin{equation}
a = k_a\, m_{q}^{\alpha}\, m_{\ell}^{\beta} \;\;\text{and}\;\; b = k_b\, m_{q}^{\gamma}\, m_{\ell}^{\delta} \;.
\end{equation}

This expresses $a$ and $b$ as power-law functions of the quark and lepton masses, fitted via linear regression in logarithmic space.

\begin{figure}[H]
    \centering
    \includegraphics[width=0.9\linewidth]{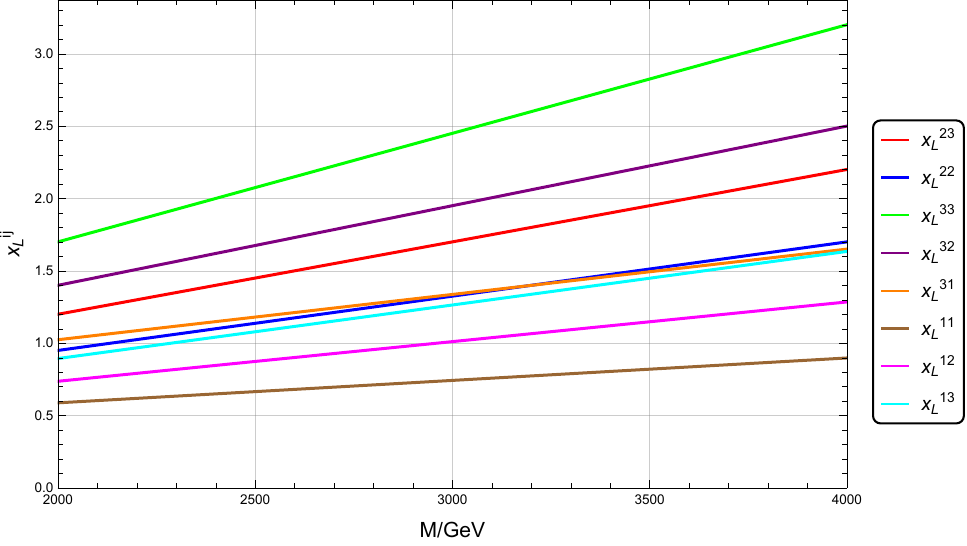}
    \caption{Upper limits of the left-handed quark-lepton couplings in $U_1$ VLQ model.}
    \label{fig:VLQCouplings}
\end{figure}

\begin{figure}[H]
    \centering
    \includegraphics[width=0.9\linewidth]{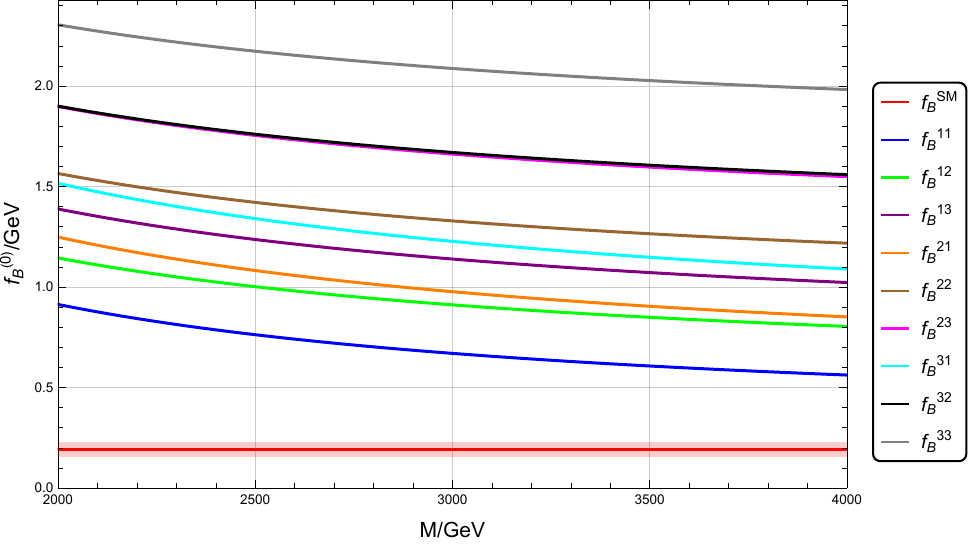}
    \caption{Effective decay constant plots from the VLQ-mediated $B$ decay at tree level, with different lepton--antineutrino combinations, using original NP couplings. With these couplings, the NP contributions to the decay constant far outweigh those from the SM.}
    \label{fig:fBtree}
\end{figure}

\begin{figure}[H]
    \centering
    \includegraphics[width=0.9\linewidth]{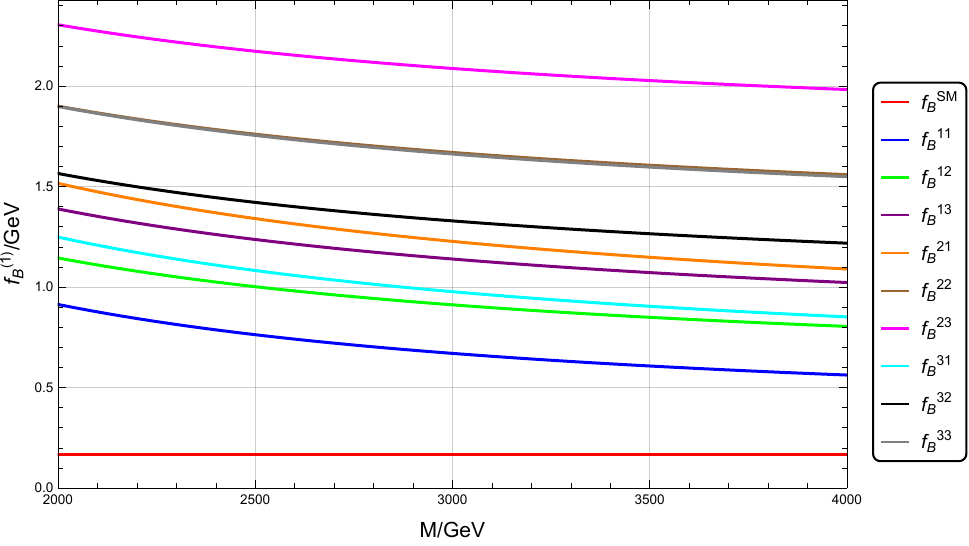}
    \caption{Effective decay constant plots from the VLQ-mediated $B$ decay at one-loop level in QCD, using original NP couplings.}
    \label{fig:fBoneloop}
\end{figure}

\begin{figure}[H]
    \centering
    \includegraphics[width=0.9\linewidth]{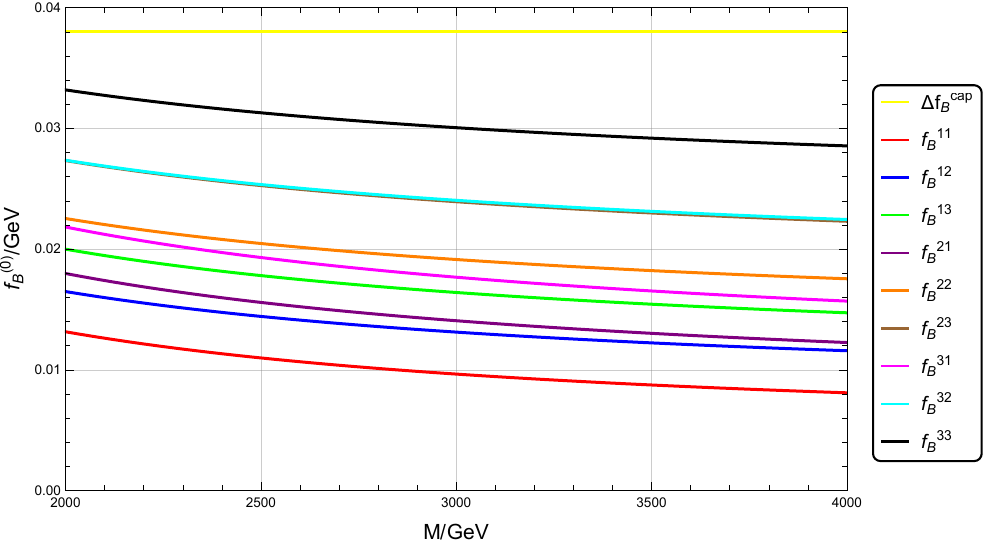}
    \caption{Effective decay constant plots from the $U_1$ VLQ-mediated $B$ decay at tree-level, using constrained NP couplings. All NP contributions fall within the upper limit cap on the uncertainty of the decay constant from the SM.}
    \label{fig:fBtreeConstrained}
\end{figure}

Fig.\ref{fig:VLQCouplings} shows the upper limits of the couplings taken from \cite{Angelescu:2021lln}, with the couplings involving light fermions obtained by power law regression. The couplings show a trend of increasing nearly linearly with leptoquark mass and roughly maintain a trend that heavy/higher generation fermion couplings are significantly larger than those between lighter fermions. This, however, is not a strict trend and there is clear overlap observed around 3\:TeV for heavy--light fermion couplings.
\bigbreak\noindent
The different plots in Figs.\ref{fig:fBtree}, \ref{fig:fBoneloop} and \ref{fig:fBtreeConstrained} correspond to the effective decay constants from amplitudes with different lepton-antineutrino combinations. The behaviour of all decay constant plots exhibits a similar trend; a gradual decrease with increasing leptoquark mass. Additionally, they preserve their hierarchical ordering according to the masses of the fermions produced, with no overlap observed across the examined range of Leptoquark masses. Consequently, this results in the effective decay constants being dominated by processes producing higher generation leptons.
\\
The QCD-improved plots in Fig.\ref{fig:fBoneloop} however, display a somewhat interesting effect in the form of a slight re-ordering of the hierarchy, most evident in amplitudes with second and third generation leptons. While this does slightly contradict the trend seen in $f_{B}^{eff}$ at tree level, it does not affect first generation leptons.

\begin{figure}[H]
    \centering
    \includegraphics[width=0.9\linewidth]{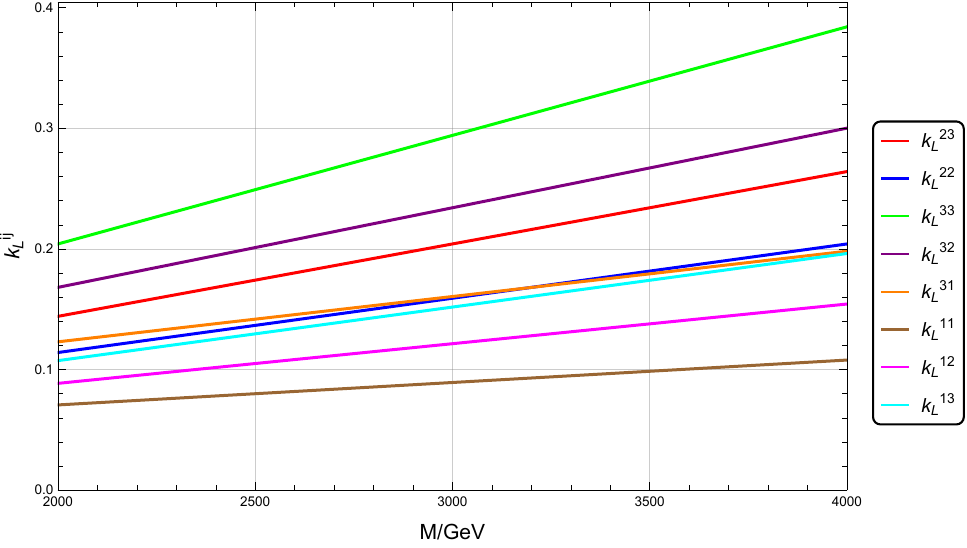}
    \caption{Scaled down upper limits of $U_1$ VLQ couplings, constrained by the uncertainty of the decay constant.}
    \label{fig:vLQCouplingsConstrained}
\end{figure}

Figs.\ref{fig:fBtree} and \ref{fig:fBoneloop} are presented as evidence that with the given upper limits of the couplings, the NP effective decay constants far exceed the limits of uncertainty. Hence we present the scaled down upper limits of these couplings in Fig.\ref{fig:vLQCouplingsConstrained}, which bring all NP tree-level contributions (Fig.\ref{fig:fBtreeConstrained}) within the bounds of uncertainty. This requires a reasonably strict and safe rescaling of the plots of Fig.\ref{fig:VLQCouplings} by $12\:\%$ of the original values.

\pagebreak

\subsection*{Branching Fractions} 

In this section we plot the upper limits of the branching fractions from NP channels, using the restricted parameter space of the $U_1$ VLQ model. These include additional contributions to the SM-compatible non-LFUV as well as the LFUV channels.
\bigbreak\noindent
The leptoquark channel branching fraction for $B\to \ell_i\,\bar{\nu}_{\ell_j}$ is given as the product of its partial decay width with the lifetime of the charged $B$ meson \cite{banerjee2024averagesbhadronchadrontaulepton}.
$$\mathcal{B}^{i\,j}=\frac{G_{F}^{2}}{8\,\pi}\,(f_{B}^{i\,j})^2\,|V_{ub}|^2\,m_B\,m_{\ell_i}^{2}\,\left(1-\frac{m_{\ell_i}^{2}}{m_{B}^{2}}\right)^2\times\tau_B$$

\begin{figure}[H]
    \centering
    \includegraphics[width=0.9\linewidth]{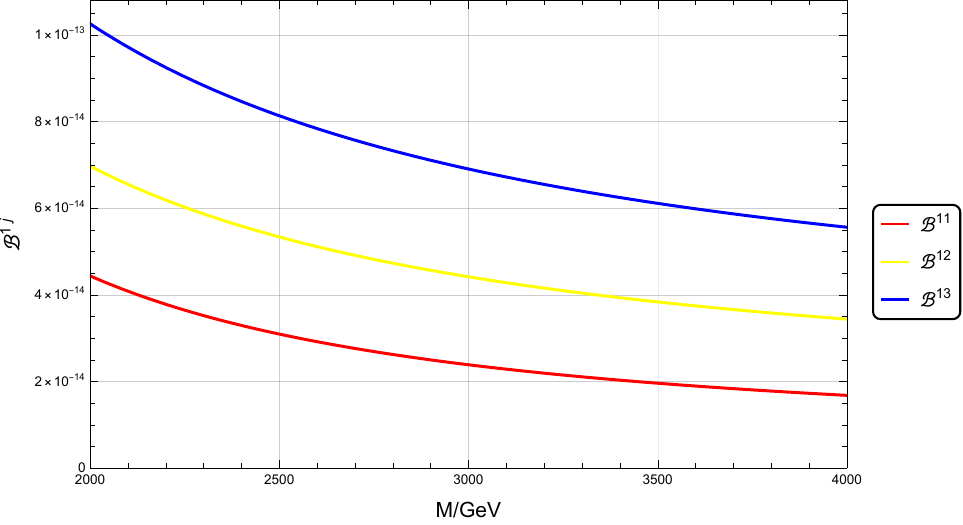}
    \caption{Branching fractions for $B\to e\,\bar{\nu_j}$ plotted against leptoquark mass.}
    \label{fig:BranchingFraction1}
\end{figure}

\begin{figure}[H]
    \centering
    \includegraphics[width=0.9\linewidth]{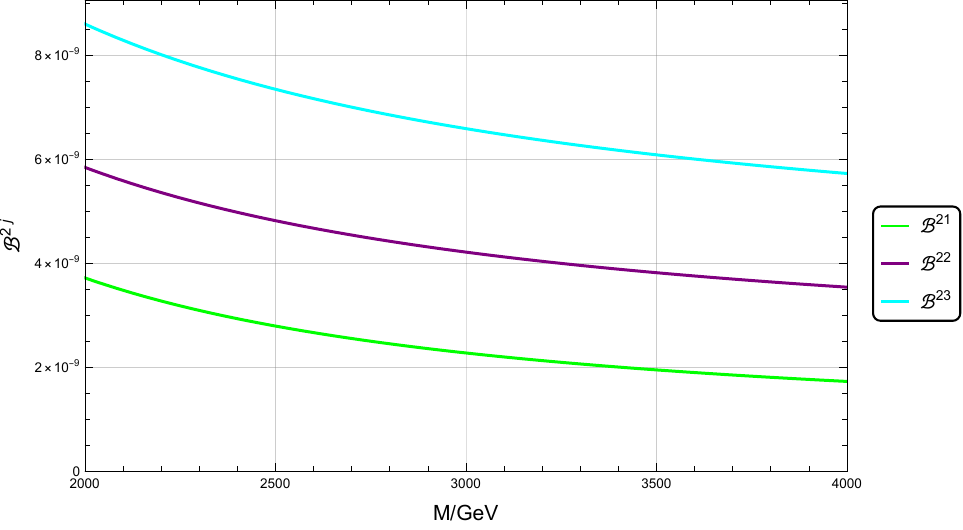}
    \caption{Branching fractions for $B\to \mu\,\bar{\nu_j}$ plotted against leptoquark mass.}
    \label{fig:BranchingFraction2}
\end{figure}

\begin{figure}[H]
    \centering
    \includegraphics[width=0.9\linewidth]{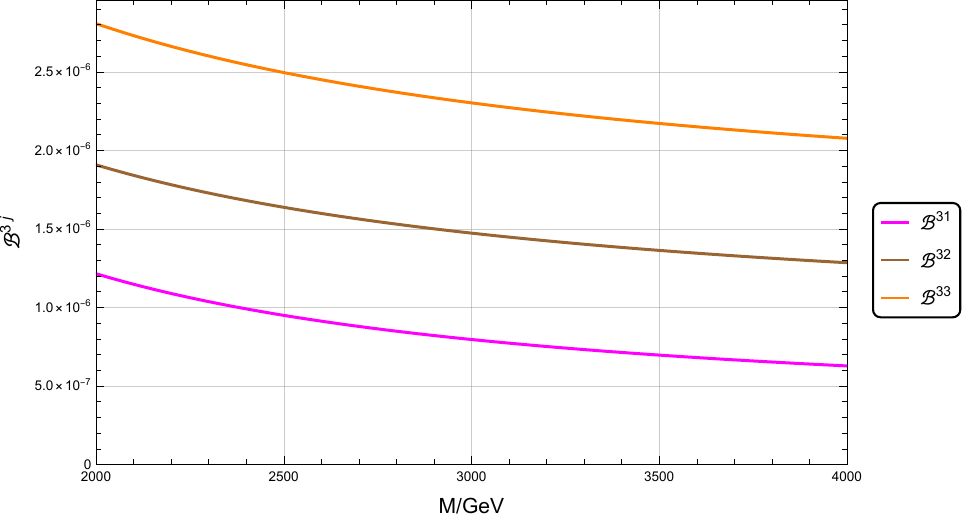}
    \caption{Branching fractions for $B\to \tau\,\bar{\nu_j}$ plotted against leptoquark mass.}
    \label{fig:BranchingFraction3}
\end{figure}

Due to the significant differences in order of magnitude, the branching fractions for the different types of charged leptons have been plotted separately (Figs.\ref{fig:BranchingFraction1},\ref{fig:BranchingFraction2} and \ref{fig:BranchingFraction3}), with the scale of magnitude primarily being determined by the charged lepton flavour. It can be seen that the non-LFUV channels are not specifically preferred, but rather the partial decay widths depend upon the lepton masses, which also dictate the couplings. As such, channels with heavier flavours are expected to contribute the most. All contributions stay well within the upper experimental bounds of the branching fractions \cite{gaudino2025mathcalbbrightarrowtaunumeasurementhadronicfei,Belle:2014pdk}, with the electron modes having a branching fraction of the order $10^{-13}-10^{-14}$, the muon modes of the order $10^{-9}$ and the tau modes of the order $10^{-6}-10^{-7}$, for $M=2-4\:$TeV.

\subsection*{Alternative New Physics Scenario}
\begin{figure}[H]
    \centering
    \includegraphics[width=0.9\linewidth]{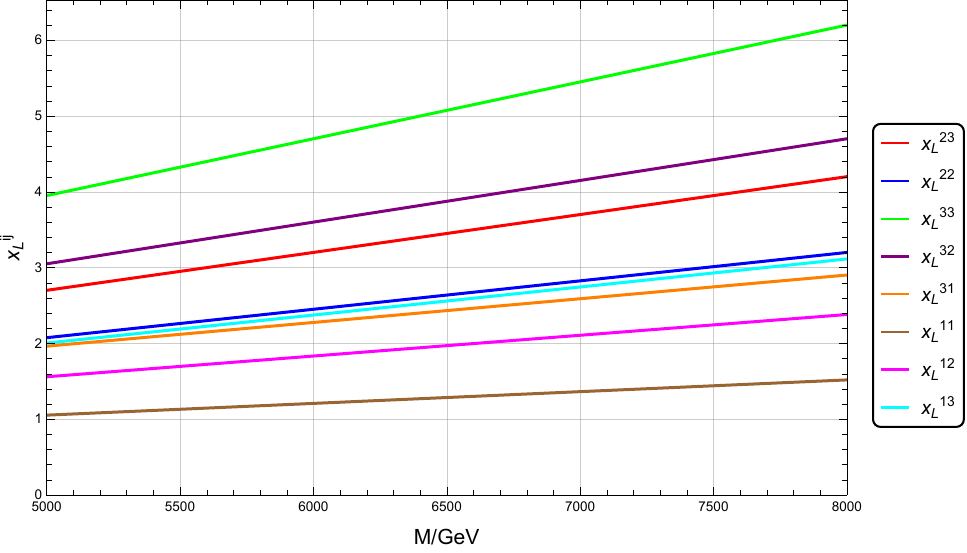}
    \caption{Quark--lepton couplings in the original model with larger leptoquark masses.}
    \label{fig:S2VLQCouplings}
\end{figure}

\begin{figure}[H]
    \centering
    \includegraphics[width=0.9\linewidth]{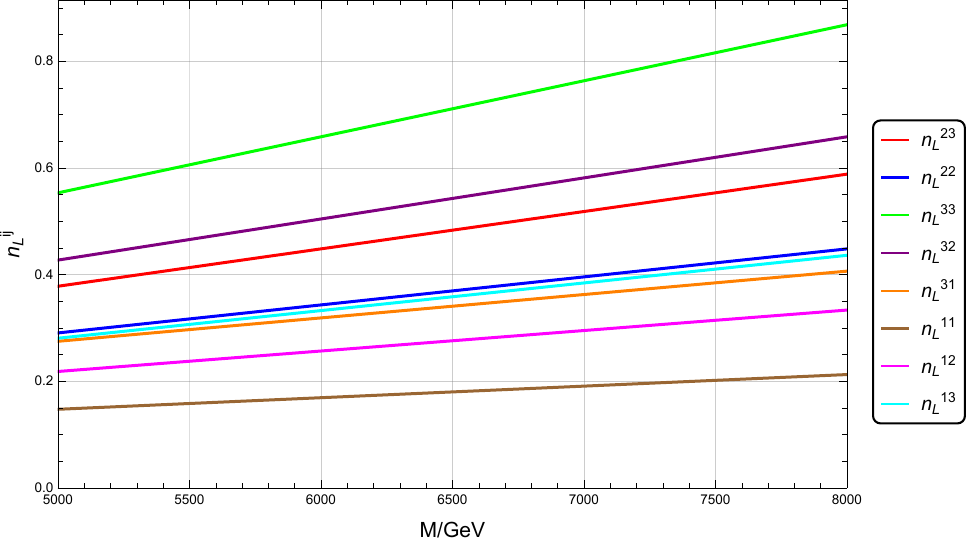}
    \caption{Constrained quark--lepton couplings with larger leptoquark masses.}
    \label{fig:S2VLQVCouplingsConstrained}
\end{figure}

\begin{figure}[H]
    \centering
    \includegraphics[width=0.9\linewidth]{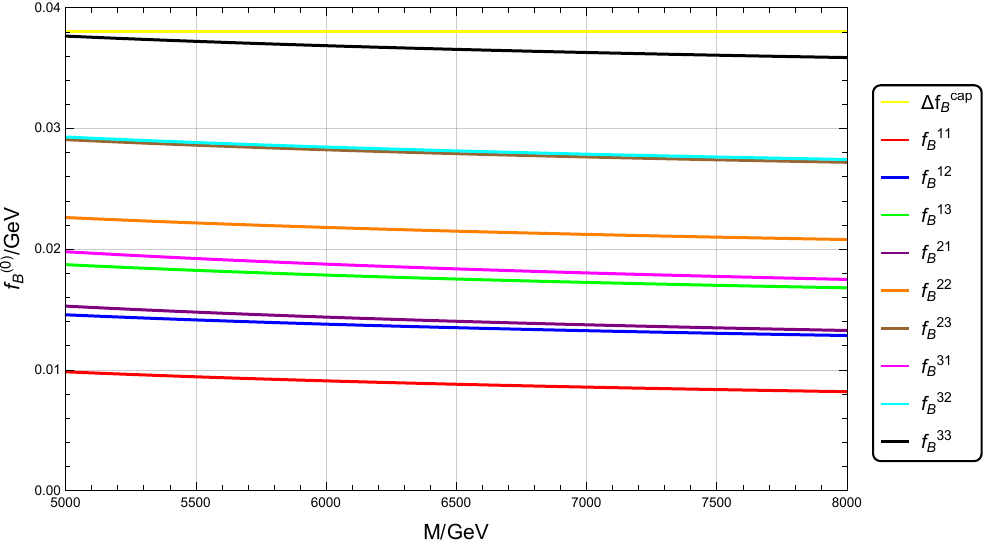}
    \caption{Effective decay constant plots from the $U_1$ VLQ-mediated $B$ decay at tree-level, using constrained NP couplings from the alternative scenario with with larger leptoquark masses.}
    \label{fig:S2fBtreeConstrained}
\end{figure}

In this section we investigate the couplings and decay constants in a different scenario of the leptoquark model. Considering a leptoquark mass range of $5-8\:$TeV, we give alternate plots for the couplings and decay constants. With larger masses, the NP decay constant contributions decrease faster hence allowing for a smaller reduction in the couplings. 
\bigbreak\noindent
In this scenario we assumed that the trend of the upper limits of the couplings in Fig.\ref{fig:VLQCouplings} holds for $M=5-8\:$TeV as well. As such, in order to constrain the NP decay constant contributions within the lattice uncertainty bounds (Fig.\ref{fig:S2fBtreeConstrained}), the strictest possible rescaling of the plots of Fig.\ref{fig:S2VLQCouplings} is by $14\:\%$ of the original values, as given in Fig.\ref{fig:S2VLQVCouplingsConstrained}.

\section*{Conclusion}

The $U_1$ Vector Leptoquark Model predicts LFU-violating amplitudes dominated by higher generation fermions. However, under the current upper bound constraints on the VLQ couplings \cite{Angelescu:2021lln}, these amplitudes lead to contributions to the $B$-meson decay constant that exceed acceptable limits. To reconcile the NP contributions with the uncertainty range of the lattice predictions for the decay constant, it is necessary to reduce the upper limits on the VLQ couplings by a substantial common scaling factor. Incorporating one-loop QCD corrections does not qualitatively alter this conclusion, although it shuffles the hierarchy of the new physics effects associated with higher generation fermions in $B$-meson decays. Alternatively, a slightly smaller reduction in the couplings can be accommodated if the allowed leptoquark mass range is extended by a few TeV. \\
The branching fractions of the leptoquark-mediated decay for non-LFUV modes lie within the experimental limits and the LFUV modes are of a similar order of magnitude. The branching fractions decrease gradually with increasing leptoquark mass but the most significant jumps in the magnitude occur with lepton flavour.
\\
The approach for this simple process and the redefined couplings can also be applied to more complicated semi-leptonic and radiative meson decays. These can be used to improve the decay form factors and study potential signatures of LFUV in the form factors and branching fractions of these more extensively studied decays.

\section*{Acknowledgments}
We extend our gratitude to S.Ishaaq and F.Bhutta for helpful feedback and discussions and to F.Ahmad for vital tech support.

\appendix

\section{Coefficients of Dirac Structures}

Following are the coefficients of the Dirac structures resulting from one-loop integration.
\bigbreak\noindent
$a_i$ are the coefficients of the eight different Dirac structures due to the vertex loop on the $b$-quark and internal leptoquark, and $b_i$ are the coefficients corresponding to a similar vertex loop with the $\bar{u}$-quark.

\begin{minipage}{0.45\textwidth}
\begin{equation}
\begin{aligned}
a_{1} &= -\tfrac{8}{3}\,m_{b}\,\mathbb{I} \;, \\
a_{2} &= -\tfrac{68}{27}\,\mathbb{I} \;-\;\tfrac{4}{9}\,\mathbb{I}\,\ln\left(\frac{\mu^{2}}{m_{b}^{2}}\right) \;, \\
a_{3} &= \tfrac{16}{3}\,m_{b}\,\mathbb{I} \;+\;\tfrac{8}{3}\,m_{b}\,\mathbb{I}\,\ln\left(\frac{\mu^{2}}{m_{b}^{2}}\right) \;, \\
a_{4} &= 4\,\mathbb{I} \;+\;\tfrac{4}{3}\,\mathbb{I}\,\ln\left(\frac{\mu^{2}}{m_{b}^{2}}\right) \;, \\
a_{5} &= \tfrac{17}{27}\,m_{b}^{2} \;+\;\tfrac{1}{9}\,m_{b}^{2}\,\ln\left(\frac{\mu^{2}}{m_{b}^{2}}\right) \;, \\
a_{6} &= \tfrac{2}{3}\,m_{b} \;, \\
a_{7} &= -\tfrac{4}{3}\,m_{b} \;-\;\tfrac{2}{3}\,m_{b}\,\ln\left(\frac{\mu^{2}}{m_{b}^{2}}\right) \;, \\
a_{8} &= -1 \;-\;\tfrac{1}{3}\,\ln\left(\frac{\mu^{2}}{m_{b}^{2}}\right) \;.
\end{aligned}
\end{equation}
\end{minipage}
\hfill
\begin{minipage}{0.45\textwidth}
\begin{equation}
\begin{aligned}
b_{1} &= -\frac{16}{3}\,m\,\mathbb{I} \;-\; \frac{8}{3}\,m\,\mathbb{I}\,\ln\!\left(\frac{\mu^{2}}{m^{2}}\right), \\
b_{2} &= 4\,\mathbb{I} \;+\; \frac{4}{3}\,\mathbb{I}\,\ln\!\left(\frac{\mu^{2}}{m^{2}}\right), \\
b_{3} &= \frac{16}{3}\,m\,\mathbb{I} \;+\; \frac{8}{3}\,m\,\mathbb{I}\,\ln\!\left(\frac{\mu^{2}}{m^{2}}\right), \\
b_{4} &= -4\,\mathbb{I} \;-\; \frac{4}{3}\,\mathbb{I}\,\ln\!\left(\frac{\mu^{2}}{m^{2}}\right), \\[4pt]
b_{5} &= \frac{4}{3}\,m \;+\; \frac{2}{3}\,m\,\ln\!\left(\frac{\mu^{2}}{m^{2}}\right), \\
b_{6} &= -1 \;-\; \frac{1}{3}\,\ln\!\left(\frac{\mu^{2}}{m^{2}}\right), \\
b_{7} &= -\frac{4}{3}\,m \;-\; \frac{2}{3}\,m\,\ln\!\left(\frac{\mu^{2}}{m^{2}}\right), \\
b_{8} &= 1 \;+\; \frac{1}{3}\,\ln\!\left(\frac{\mu^{2}}{m^{2}}\right).
\end{aligned}
\end{equation}
\end{minipage}

$c_i$ are the coefficients corresponding to the Dirac structures of the hadronic box loop.

\begin{equation}
\begin{aligned}
c_{1} &= -4 \;-\; 2\,\ln\!\left(\frac{m^{2}}{m_{b}^{2}}\right) \;, \\
c_{2} &= -\frac{2}{m_{b}} \;-\; \frac{2}{m_{b}}\,\ln\!\left(\frac{m^{2}}{m_{b}^{2}}\right) \;.
\end{aligned}
\end{equation}

\section{Hadronic Matrix Elements}

Here we derive the parameterization of the different quark bilinears appearing in the amplitudes.
\bigbreak\noindent
As a starting point we consider the tree-level matrix element and the axial-vector current divergence given below. From these, other matrix elements can be derived.

\begin{equation}
\boxed{
    -\langle 0 | \bar{u}\,\gamma^{\mu}\,P_L\,b | B \rangle= \langle 0 | \bar{u}\,\gamma^{\mu}\,\gamma^5\,b | B \rangle =i\,f_B\,p^{\mu} \;, 
    \label{eq:PL_gamma_mu}
    }
\end{equation}

\begin{equation}
    \partial_{\mu}\,(\bar{u}\,\gamma^{\mu}\,\gamma^5\,b)=i\,(m_b+m)\,\bar{u}\,\gamma^5\,b \;.
    \label{eq:Axial_vector_div}
\end{equation}

Using \eqref{eq:Axial_vector_div} within \eqref{eq:PL_gamma_mu} (the vector bilinear does not contribute),

\begin{equation}
   \langle 0 |\partial_{\mu}\,(\bar{u}\,\gamma^{\mu}\,\gamma^5\,b) | B \rangle =\langle 0 |i\,p_{\mu}\,(\bar{u}\,\gamma^{\mu}\,\gamma^5\,b) | B \rangle =i\,(m_b+m)\,\langle 0 |\bar{u}\,\gamma^5\,b | B \rangle \;,
    \label{eq:PL_gamma-nu_1}
\end{equation}

contracting \eqref{eq:PL_gamma_mu} through with $p_{\mu}$

\begin{equation}
    p_{\mu}\langle 0 | \bar{u}\,\gamma^{\mu}\,\gamma^5\,b | B \rangle =i\,f_B\,m_{B}^{2} \;,
\end{equation}

Substituting it into \eqref{eq:PL_gamma-nu_1} gives a very useful intermediate result.

\begin{equation}
    -f_B\,m_{B}^{2}=i\,(m_b+m)\,\langle 0 |\bar{u}\,\gamma^5\,b | B \rangle \implies \langle 0 |\bar{u}\,\gamma^5\,b | B \rangle = i\,f_B\,\frac{m_{B}^{2}}{m_b+m} \;.
    \label{eq:gamma5}
\end{equation}

Using this result to derive a matrix element:

\begin{equation}
    \langle 0 | \bar{u}\,P_L\,p^{\mu}\,b | B \rangle=p^{\mu}\, \langle 0 | \bar{u}\,b | B \rangle - p^{\mu}\, \langle 0 | \bar{u}\,\gamma^5\,b | B \rangle \;.
\end{equation}

As the matrix element of the scalar bilinear is zero so

\begin{equation}
\boxed{
    \langle 0 | \bar{u}\,P_L\,p^{\mu}\,b | B \rangle=-i\,f_B\,\frac{m_{B}^{2}}{m_b+m}\,p^{\mu} \;.
    \label{eq:PL_p_mu}
    }
\end{equation}

$\langle 0 | \bar{u}\,P_L\,\slashed{p}\,p^{\mu}\,b | B \rangle$, can be expanded as follows, with the vector part not contributing. Substituting \eqref{eq:PL_gamma_mu},

\begin{equation}
    \langle 0 | \bar{u}\,P_L\,\slashed{p}\,p^{\mu}\,b | B \rangle=-p^{\mu}\,p_{\nu}\,\langle 0 | \bar{u}\,\gamma^{\nu}\,\gamma^5\,b | B \rangle=-i\,f_B\,m_{B}^{2}\,p^{\mu}\;.
\end{equation}

Using a similar approach as above but with the $B$-meson momentum replaced by external lepton momentum, 

\begin{equation}
\boxed{
\langle 0 | \bar{u}\,P_L\,\slashed{p}\,p_{\ell}^{\mu}\,b | B \rangle= -i\,f_B\,m_{B}^{2}\,p_{\ell}^{\mu} \;, \label{eq:PL_pslash_p_l_mu}
}
\end{equation}

and 

\begin{equation}
\boxed{
\langle 0 | \bar{u}\,P_L\,p_{\ell}^{\mu}\,b | B \rangle= -i\,f_B\,\frac{m_{B}^{2}}{m_b+m_u}\,p_{\ell}^{\mu} \;. 
\label{eq:PL_p_l_mu} 
}
\end{equation}

$\langle 0 | \bar{u}\,P_L\,\slashed{p}\,\gamma^{\mu}\,b | B \rangle$ can be derived  by expressing it as a vector with a scalar coefficient,

\begin{equation}
    \langle 0 | \bar{u}\,P_L\,\slashed{p}\,\gamma^{\mu}\,b | B \rangle=F\,p^{\mu} \;.
    \label{eq:PL_pslash_gamma_mu}
\end{equation}

Contracting both sides of \eqref{eq:PL_pslash_gamma_mu} with $p^{\mu}$ and with only the axial-vector bilinear contributing,

\begin{equation}
    p^{\mu}\,\langle 0 | \bar{u}\,P_L\,\slashed{p}\,\gamma^{\mu}\,b | B \rangle=\langle 0 | \bar{u}\,P_L\,\slashed{p}^2\,b | B \rangle=F\,m_{B}^{2} \implies F=-\frac{1}{2}\, \langle 0 | \bar{u}\,\gamma^5\,b | B \rangle\;.
\end{equation}

Using the earlier result of \eqref{eq:gamma5},

\begin{equation}
    \boxed{\langle 0 | \bar{u}\,P_L\,\slashed{p}\,\gamma^{\mu}\,b | B \rangle=-i\,f_B\,\frac{m_{B}^{2}}{m_b+m}\,p^{\mu}\;.
    \label{eq:PL_pslash_gamma-mu}
    }
\end{equation}

$\langle 0 | \bar{u}\,P_L\,\slashed{p}_{\ell}\,\gamma^{\mu}\,b | B \rangle$ is similar to the above but with the $B$-meson momentum swapped with the lepton momentum, hence

\begin{equation}
\boxed{
  \langle 0 | \bar{u}\,P_L\,\slashed{p}_{\ell}\,\gamma^{\mu}\,b | B \rangle= -i\,f_B\,\frac{m_{B}^{2}}{m_b+m}\,p_{\ell}^{\mu} \;.
  \label{eq:PL_pslashl_gamma_mu} 
  }
\end{equation}

$\langle 0 | \bar{u}\, P_{L} \,\slashed{p}\,\slashed{p}_{\ell}\,\gamma^{\mu} \,b | B(p) \rangle$ can be derived by making use of \eqref{eq:PL_gamma_mu} and a gamma matrices identity,

\begin{equation}
    \slashed{p}\,\slashed{p}_{\ell}\,\gamma^{\mu}= (p\!\cdot\!p_{\ell})\:\gamma^{\mu}
- p^{\mu}\,\slashed{p}_{\ell}
+ p_{\ell}^{\mu}\,\slashed{p}
+ i\,\epsilon^{\alpha\beta\mu\nu}\,p_{\alpha}\,p_{{\ell}\,\beta}\,\gamma_{\nu}\,\gamma^{5}.
\label{eq:gamma_identity}
\end{equation}

The Levi-Cevita term vanishes for a pseudoscalar, leaving

\begin{equation}
\langle 0 | \bar{u}\, P_{L} \,\slashed{p}\,\slashed{p}_{\ell}\,\gamma^{\mu} \,b | B \rangle
= (p\!\cdot\!p_{\ell})\:\langle 0 | \bar{u}\, P_{L}\,\gamma^{\mu} \,b | B \rangle
- p^{\mu}\:\langle 0 | \bar{u} \,P_{L}\,\slashed{p}_{\ell} \,b | B \rangle + p_{\ell}^{\mu}\:\langle 0 | \bar{u} \,P_{L}\,\slashed{p} \,b | B \rangle \;.
\end{equation}

Using the earlier result \eqref{eq:PL_gamma_mu} and two further intermediate results:

\begin{equation}
    \langle 0 | \bar{u} \, P_{L} \,\slashed{p}_{\ell}  \, b | B \rangle = -i\,f_{B} \:(p\!\cdot\!p_{\ell}) \;,
\end{equation}

and 

\begin{equation}
    \langle 0 | \bar{u} \, P_{L} \,\slashed{p} \, b | B \rangle = -i\,f_{B} \,m_{B}^{2} \;.
\end{equation}

The first two terms cancel, leaving

\begin{equation}
\boxed{
\langle 0 | \bar{u} \,P_{L} \,\slashed{p}\,\slashed{p}_{\ell}\,\gamma^{\mu}\, b | B(p) \rangle
= -i f_{B}\,m_{B}^{2}\,p_{\ell}^{\mu} \;.
\label{eq:PL_pslash_plslash_gamma-mu}
}
\end{equation}

For deriving the matrix element $\langle 0 | \bar{u}\,P_L\,\slashed{k}\,p^{\mu}\,b | B \rangle$, start off with the up quark equation of motion $i\slashed{D}\,u=m\,u$ and approximate it inside the meson as $i\,\slashed{D}\,u=\slashed{k}\,u$. Comparing the two gives $\slashed{k}\,u=m\,u$. Using this result inside the following LHS matrix element gives

\begin{equation}
    \langle 0 | \bar{u}\,P_L\,\slashed{k}\,p^{\mu}\,b | B \rangle=m\,p^{\mu}\,\langle 0 | \bar{u}\,P_L\,b | B \rangle
\end{equation}

Invoking the earlier result of \eqref{eq:gamma5},

\begin{equation}
    \boxed{\langle 0 | \bar{u}\,P_L\,\slashed{k}\,p^{\mu}\,b | B \rangle=-i\,f_B\,\frac{m\,m_{B}^{2}}{m_b+m}\,p^{\mu} \;.
    \label{eq:PL_kslash_p_mu}
    }
\end{equation}

To derive $\langle 0 | \bar{u}\,P_L\,\slashed{k}\,\slashed{p}\,\gamma^{\mu}\,b | B \rangle$, use the up quark equation of motion in the matrix element as follows

\begin{equation}
    \langle 0 | \bar{u}\,P_L\,\slashed{k}\,\slashed{p}\,\gamma^{\mu}\,b | B \rangle=m\,\langle 0 | \bar{u}\,P_L\,\slashed{p}\,\gamma^{\mu}\,b | B \rangle \;.
\end{equation}

Invoking the earlier result of \eqref{eq:PL_pslash_gamma-mu},

\begin{equation}
\boxed{
\langle 0 | \bar{u}\,P_L\,\slashed{k}\,\slashed{p}\,\gamma^{\mu}\,b | B \rangle= -i\,f_B\,\frac{m\,m_{B}^{2}}{m_b+m}\,p^{\mu} \;,
}
\end{equation}

Using similar approaches as \eqref{eq:PL_kslash_p_l_mu} and \eqref{eq:PL_kslash_p_mu} but with lepton momentum, gives the two matrix elements below

\begin{equation}
\boxed{
\langle 0 | \bar{u}\,P_L\,\slashed{k}\,p_{\ell}^{\mu}\,b | B \rangle
= -i\,f_B\,\frac{m\,m_{B}^{2}}{m_b+m}\,p_{\ell}^{\mu} \;.
\label{eq:PL_kslash_p_l_mu}
}
\end{equation}

and

\begin{equation}
\boxed{
\langle 0 | \bar{u}\,P_L\,\slashed{k}\,\slashed{p}_{\ell}\,\gamma^{\mu}\,b | B \rangle= -i\,f_B\,\frac{m\,m_{B}^{2}}{m_b+m}\,p_{\ell}^{\mu} \;.
}
\end{equation}

\bibliographystyle{ieeetr}
\bibliography{main.bib}

\end{document}